\documentclass[article,nojss]{jss} 
\pdfoutput=1
\usepackage{graphics}
\usepackage{epsf}
\usepackage{amssymb}
\usepackage{amsmath}
\usepackage{amsthm}
\usepackage{latexsym}
\usepackage{natbib}
\usepackage{bm}
\usepackage{color}
\usepackage[T1]{fontenc}
\theoremstyle{definition}

\theoremstyle{plain}

\newcommand{\beq}{\begin{equation}}
\newcommand{\eeq}{\end{equation}}

\newcommand{\ts}{^\top}

\newcommand{\bmat}[1]{\left [ \begin{array}{#1}}
\newcommand{\emat}{\end{array}\right ]}

\title{Just Another Gibbs Additive Modeller: Interfacing \proglang{JAGS} and \pkg{mgcv}}
\Plaintitle{Just Another Gibbs Additive Modeller: Interfacing JAGS and mgcv}
\Shorttitle{Just Another Gibbs Additive Modeller}
\author{Simon N. Wood\\University of Bristol}
\Plainauthor{Simon N. Wood}
\Address{Simon N. Wood,\\School of Mathematics\\
University of Bristol, BS8 1TW U.K.\\
E-mail: \email{simon.wood@r-project.org}
}

\Abstract{
The \proglang{BUGS} language offers a very flexible way of specifying complex statistical models for the purposes of Gibbs sampling, while its \proglang{JAGS} variant offers very convenient \proglang{R} integration via the \pkg{rjags} package. However, including smoothers in \proglang{JAGS} models can involve some quite tedious coding, especially for multivariate or adaptive smoothers. Further, if an additive smooth structure is required then some care is needed, in order to centre smooths appropriately, and to find appropriate starting values. \proglang{R} package \pkg{mgcv} implements a wide range of smoothers, all in a manner appropriate for inclusion in \proglang{JAGS} code, and automates centring and other smooth setup tasks. The purpose of this note is to describe an interface between \pkg{mgcv} and \proglang{JAGS}, based around an \proglang{R} function, \code{jagam}, which takes a generalized additive model (GAM) as specified in \pkg{mgcv} and automatically generates the \proglang{JAGS} model code and data required for inference about the model via Gibbs sampling. Although the auto-generated \proglang{JAGS} code can be run as is, the expectation is that the user would wish to modify it in order to add complex stochastic model components readily specified in \proglang{JAGS}. A simple interface is also provided for visualisation and further inference about the estimated smooth components using standard \pkg{mgcv} functionality. The methods described here will be un-necessarily inefficient if all that is required is fully Bayesian inference about a standard GAM, rather than the full flexibility of \proglang{JAGS}. In that case the \pkg{BayesX} package would be  more efficient.
}
\Plainkeywords{R, BUGS, JAGS, additive model, spline, smooth, generalized additive mixed model}
\begin{document}

\section{Introduction}

\begin{figure} 
\centering
\includegraphics[angle=-90,scale=0.6]{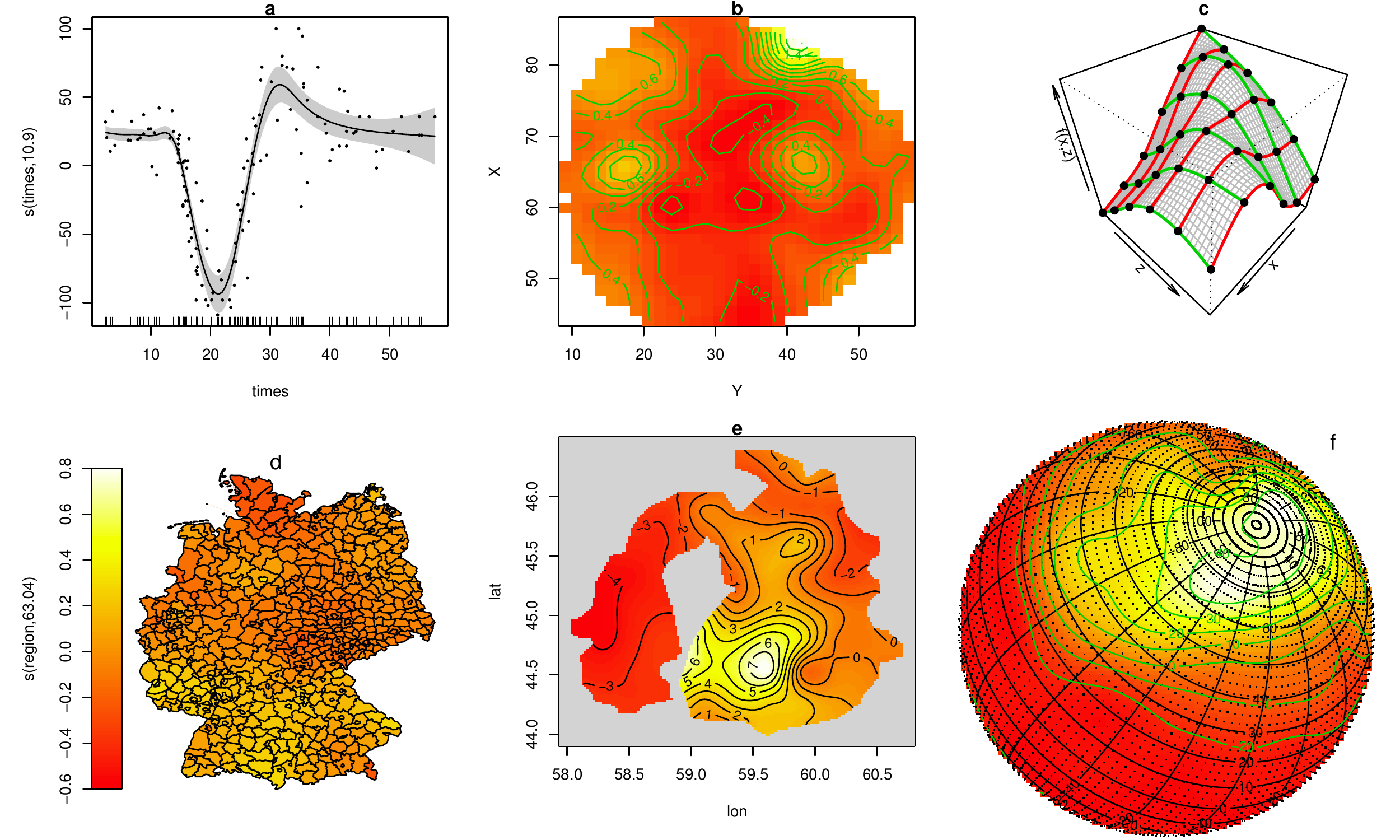}
\caption{Some of the rich variety of smooths available in the \pkg{mgcv} package. From top right: A simple one dimensional adaptive smooth; multidimensional thin-plate splines; multidimensional tensor product smooths; Gaussian Markov random fields; soap film finite area smoothers; splines on the sphere. \label{smooths}}
\end{figure}

This paper is about automatically and reliably generating \proglang{JAGS} \citep{plummer2003jags} model specification code and data implementing any generalized additive model \citep[GAM,][]{h&t90} that can be specified in the \proglang{R} \citep{rcore} package \pkg{mgcv} \citep{wood2006igam,mgcv}. The purpose of this is to allow models with the complex smooth structure permitted by \pkg{mgcv} (exemplified by Figure \ref{smooths}) combined with the complex random structure permitted by \proglang{JAGS} to be produced more easily than has hitherto been the case. As the paper's title makes clear, there is nothing new about using Markov chain Monte Carlo (MCMC) in general, or Gibbs sampling in particular, for smooth modelling. The paper's purpose is simply to make this easier and more automatic and hence less susceptible to implementation error, and to document the methods used to achieve this.

In principle, the \proglang{JAGS} package and language allows Bayesian inference about a very wide range of models that can be written as directed acyclic graphs (DAG). This class includes GAMs as one special case. The Bayesian view of spline smoothing and additive models is almost as old as splines and additive models themselves \citep{kimeldorfwahba1970, wahba83, silverman85, hastie2000bayesBF, fahrmeir.lang}, and several authors have exploited this to use \proglang{JAGS} or \proglang{BUGS} \citep{bugs} for generalized additive modelling, notably \citet{crainiceanu2005} based on \citet{ruppert.wand.carroll} and \citet{zuurGAMM}.

In principle the \pkg{mgcv} package already included all the code required to set up smoothers for use with \proglang{JAGS}. This is because what is required is essentially the same as what is required to use any standard mixed modelling software for GAM inference: for example \pkg{mgcv} function \code{gamm} based on the appendix of \citet{wood04} uses the \pkg{nlme} package \citep{nlme} in this way. However, a considerable degree of user expertise is required to implement this reliably in practice. 

A particular area where difficulty can arise is in the use of centring constraints on model smooth components. Usually additive smooth model structures only make statistical sense if such constraints are applied \citep[see e.g.,][]{h&t90}, otherwise there is a global intercept associated with each smooth. However the \proglang{JAGS} requirement for all priors to be proper, means that failing to implement such constraints will not cause complete failure of Gibbs sampling. Instead one may see very wide credible intervals and poor mixing, but not realise that this is a model formulation problem rather than a statistical inevitability.
  
\section[The jagam function]{The \code{jagam} function}

The new \pkg{mgcv} function \code{jagam} is designed to be called in the same way that the modelling function \code{gam} would be called. That is, a model formula and family object specify the required model structure, while the required data are supplied in a data frame or list or on the search path. However, unlike \code{gam}, \code{jagam} does no model fitting. Rather it writes \proglang{JAGS} code to specify the model as a Bayesian graphical model for simulation with \proglang{JAGS}, and produces a list containing the data objects referred to in the \proglang{JAGS} code, suitable for passing to \proglang{JAGS} via the \pkg{rjags} \citep{rjags} function \code{jags.model}.

A simple model, with two univariate smooths and one tensor product smooth, exemplifies the approach. Suppose that we have a data frame, \code{dat}, containing the response and predictor variables, have loaded the \pkg{mgcv} package and have used \code{setwd} to set the working directory to something appropriate. The code
\begin{Code}
R> jd <- jagam(y ~ s(x0) + te(x1, x2) + s(x3), data = dat,
R+             family = Gamma(link=log), file = "test.jags")
\end{Code}
would specify a simple log gamma additive model structure,
$$
\log(\mu_i) = f_1(x_{0i}) + f_2(x_{1i},x_{2i}) + f_3(x_{3i}), ~~~~ y_i \sim \Gamma(\mu_i,\phi),
$$
where $f_2$ is a scale invariant tensor product smoother, appropriate for representing smooth interaction terms. \code{jagam} returns a list containing standard \pkg{mgcv} GAM setup information (\code{pregam}) and a list, \code{jags.data}, containing the objects required by \proglang{JAGS} for model simulation. The function also writes a \proglang{JAGS} model specification in the file \code{test.jags}, as follows.
\begin{Code}
model {
  eta <- X 
  for (i in 1:n) { mu[i] <-  exp(eta[i]) } ## expected response
  for (i in 1:n) { y[i] ~ dgamma(r,r/mu[i]) } ## response 
  r ~ dgamma(.05,.005) ## scale parameter prior 
  scale <- 1/r ## convert r to standard GLM scale
  ## Parameteric effect priors CHECK tau is appropriate!
  for (i in 1:1) { b[i] ~ dnorm(0,0.001) }
  ## prior for s(x0)... 
  K1 <- S1[1:9,1:9] * lambda[1]  + S1[1:9,10:18] * lambda[2]
  b[2:10] ~ dmnorm(zero[2:10],K1) 
  ## prior for te(x1,x2)... 
  K2 <- S2[1:24,1:24] * lambda[3]  + S2[1:24,25:48] * lambda[4] + 
        S2[1:24,49:72] * lambda[5]
  b[11:34] ~ dmnorm(zero[11:34],K2) 
  ## prior for s(x3)... 
  K3 <- S3[1:9,1:9] * lambda[6]  + S3[1:9,10:18] * lambda[7]
  b[35:43] ~ dmnorm(zero[35:43],K3) 
  ## smoothing parameter priors CHECK...
  for (i in 1:7) {
    lambda[i] ~ dgamma(.05,.005)
    rho[i] <- log(lambda[i])
 }
}
\end{Code}
The comments are auto-generated and designed to make it easy to locate the model components, and to draw attention to parts that the user might wish to modify. 


In normal use the file would be edited to include the more complex stochastic components likely to have been the motivation for taking a Gibbs sampling approach. It can of course be used un-modified to simply simulate from the posterior of the model parameters, as in the following example code.
\begin{Code}
R> require(rjags)
R> jm <- jags.model("test.jags", data = jd$jags.data,
R+                  inits = jd$jags.ini, n.adapt = 2000, n.chains = 1)
R> sam <- jags.samples(jm, c("b", "rho", "scale"), n.iter = 10000,
R+                     thin = 10)
\end{Code}
The chains should then be checked for convergence and reasonable mixing in the standard ways. \proglang{R} package \pkg{coda} facilitates this \citep{coda}. 

\begin{figure}
\centering
\includegraphics[angle=-90,scale=0.4]{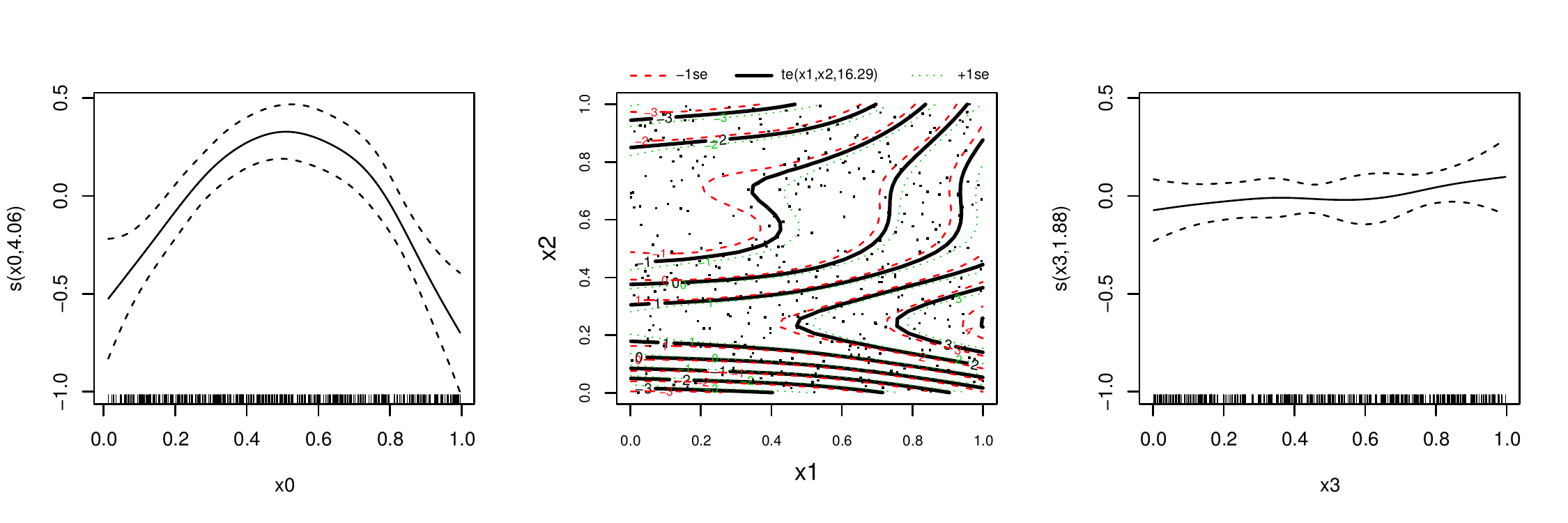}
\caption{Plots of \proglang{JAGS} estimated smooth components of a log gamma additive model. \label{gamma-eg}}
\end{figure}

If all is in order, then many users would want to use the simulation output directly, but the utility function \code{sim2gam}  can also be used to convert the simulation output into a reduced version of a fitted gam object, suitable for further use with standard \pkg{mgcv} functions. For example
\begin{Code}
R> jam <- sim2jam(sam, jd$pregam)
R> par(mfrow = c(1,3)); plot(jam)
\end{Code}
yields Figure \ref{gamma-eg}.

\section{The underlying theory}
\subsection[Smoothers in JAGS]{Smoothers in {\proglang{JAGS}} }

In \pkg{mgcv}, smooth functions are represented using spline like basis expansions, with quadratic penalties on the basis coefficients being used to avoid overfit. Generically a function $f(x)$ may be represented as 
$$
f(x) = \sum_{j=1}^K \beta_j b_j(x) 
$$
where the $\beta_j$ are unknown model parameters/coefficients, and the $b_j(x)$ are spline like basis functions. $K $ is chosen to be large enough to avoid oversmoothing, but small enough to avoid excessive computational cost. The fitted flexibility of $f$ is controlled less by $K$ than by the imposition, during fitting, of a quadratic smoothing penalty of the form
$$
\sum_j \lambda_j {\bm \beta} \ts {\bf S}_j \bm \beta,
$$
where the ${\bf S}_j$ are matrices of known coefficients and the $\lambda_j$ are smoothing parameters to be estimated. Often there is only a single component in the penalty, but the general summation form is necessary in order to implement adaptive and tensor product smooths, for example, and will also be used below to ensure the propriety of priors required for Gibbs sampling with \proglang{JAGS}. See chapter 4 of \citet{wood2006igam} for more detail. 

\citet{kimeldorfwahba1970}, \citet{wahba83} and  \citet{silverman85} provide the basis for viewing such smoothers in a Bayesian way, with the penalties induced by improper multivariate Gaussian priors, having precision matrices proportional to $\sum_j \lambda_j {\bf S}_j$. Adopting this viewpoint it is obvious that we can make inferences about smooths using Bayesian methods. 

For practical Gibbs sampling in \proglang{JAGS} there are two cases to distinguish:
\begin{enumerate}
\item those in which the prior precision matrix can be represented as a weighted sum of matrices that are all zero, apart from some unit entries on the leading diagonal, where no component matrices of the sum have unit entries in the same place.
\item those in which the precision matrix can not be written in the above form.
\end{enumerate}
Case 1 results in i.i.d. Gaussian priors on separate subsets of the parameters. For single smoothing parameter smooths it is possible to re-parameterize to achieve this case. Case 2 results in non-independent multivariate normal priors. In both cases the prior implied by the smoothing penalty is usually improper, as the penalty usually leaves some subspace of functions unpenalized. For Gibbs sampling with \proglang{JAGS} we require proper priors, but this is easily arranged. 

\subsubsection{Independent Gaussian prior smooths}

Some smooths, such as the tensor product smoothers constructed by \citet{wood2013t2} or the truncated power basis P-splines advocated by \citet{ruppert.wand.carroll}, automatically have penalties in which the ${\bf S}_j$ are identity matrices with some unit entries set to zero (and no unit entries in common between different ${\bf S}_j$). Let ${\bm \beta}_j$ denote the set of coefficients for which the corresponding diagonal elements of ${\bf S}_j$ are 1 rather than zero. Then the prior on each element of ${\bm \beta}_j$ is $N(0,1/\lambda_j)$ and the elements are independent a priori. 
A vague prior would typically be placed on $\lambda_j$. If ${\bm \beta}_0 $ denotes the coefficients that are unpenalized, we can impose a prior $N(0,1/\lambda_0)$ on these, where $\lambda_0$ is  so small as to force the prior to be very vague, or $\lambda_0$ itself has a vague prior. 

Any smooth that only has a single ${\bf S}_j$ and single $\lambda_j$ can be re-parameterized to have a partial identity matrix prior precision matrix following \citet{wood04}. Dropping the index $j$, we find the symmetric eigen-decomposition ${\bf S} = {\bf U}{\bm \Lambda}{\bf U}\ts$. Suppose that the final $M$ eigenvalues on the leading diagonal of $\bm \Lambda $ are zero (the remainder being positive). Define $\bf D$ to be the diagonal matrix with diagonal elements consisting of the square roots of the positive eigen-values from $\bm \Lambda$, followed by $M$ unit entries. If we now adopt the re-parameterization ${\bm \beta}^\prime = {\bf DU}\ts {\bm \beta}$ then the penalty matrix $\bf S$ becomes the identity matrix, with the last $M$ leading diagonal elements set to zero. The corresponding model matrix is then ${\bf XUD}^{-1}$. The last $M$ elements of ${\bm \beta}^\prime$ are now un-penalized, and, as above, this can be handled by imposing independent $N(0,1/\lambda_0)$ priors on these. The same result could be achieved somewhat more efficiently with a pivoted Cholesky decomposition. \pkg{mgcv} contains functions to perform this reparameterization automatically. Note that the preceding re-parameterization is slightly different to that employed in \citet{crainiceanu2005}, which starts from an indefinite ${\bf S}$, so that the reparameterization step also involves an element of approximation. 

\subsubsection{General Gaussian prior smooths}

For a Gaussian likelihood, independent prior smooths can result in quite fast computation, because \proglang{JAGS} is then able to employ conjugate samplers. Similarly in generalized linear model settings, the samplers from the \proglang{JAGS} \code{glm} module can also lead to efficient computation. However outside these settings the independent prior approach is slow and block updates are preferable. In any case there are several important smoother classes that are not susceptible to writing in independent prior form, notably adaptive smooths and several types of tensor product smooth. 

In fact implementing any quadratically penalized smoother in \proglang{JAGS} is straightforward, using \code{dmnorm}, the \proglang{JAGS} multivariate normal density. \code{dmnorm} is parameterized in terms of a precision matrix, for which $\sum_j \lambda_j {\bf S}_j$ can be used directly. 

The only potential difficulty is that  $\sum_j \lambda_j {\bf S}_j$ itself is usually rank deficient, implying an improper prior for the smooth. Again we must construct a prior for the null space of the smoothing penalty, but again it is possible to re-use existing \pkg{mgcv} facilities. Specifically, in the context of model selection, \citet{marra.wood2011} propose a simple construction of a penalty on the null space related to the re-parameterization used in the previous section. Again use a symmetric eigen-decomposition $\sum_j {\bf S}_j = {\bf U }{\bm \Lambda}{\bf U}\ts$. Now let ${\bf U}_0$ denote the columns of $\bf U$ (eigenvectors) corresponding to zero eigenvalues. Let ${\bf S}_0 = {\bf U}_0 {\bf U}_0\ts$. $\lambda_0 {\bm \beta}\ts {\bf S}_0 {\bm \beta} $ can be used to penalize the null space of the smoother by adding $\lambda_0{\bf S}_0$ to the precision matrix, hence making the prior on $\bm \beta$ proper. \pkg{mgcv} can generate such null space penalties automatically.

\subsubsection{Smoothing parameter priors}

\code{jagam} automates two possibilities for smoothing parameter priors: vague gamma priors on the $\lambda_j$, or bounded uniform priors on $\rho_j = \log \lambda_j$. The former will be conjugate in a fully Gaussian setting, but the latter may be considered more interpretable for the user used to thinking about log smoothing parameters. 

\subsection{Centring the smoothers}

As constructed so far, each smooth in an additive model would include its own global intercept. The data provides no information to identify these multiple intercepts, so they are only formally identifiable because of the priors put on them, which are vague priors of convenience. This lack of statistically meaningful identifiability will serve to substantially inflate credible intervals and promote slow mixing, so it is preferable to remove the redundant intercepts from the model. In an additive model context this is usually done by centring the smooths \citep{h&t86,h&t90,chambers1991,wood2006igam}. That is we impose the condition that each smooth should sum to zero over the observed values its covariates. i.e., $\sum_{i=1}^n f(x_i) = 0$. Other constraints are possible, but generally give wider credible intervals for the constrained smooths \citep[see section 4 of][for a discussion]{wood2013t2}. \pkg{mgcv} has facilities to simply absorb centring constraints into the basis by reparameterization, as described in section 4.2 of \cite{wood2006igam}. This absorption is done before any reparameterization or construction of priors on the null space. 

\subsection{Initial values}

In the Gaussian likelihood case, with gamma priors for the smoothing parameters, the initial values of the model coefficients and smoothing parameters are rather unimportant. In this situation conjugate samplers are used and, although poor starting values may prolong burn-in, eventually good results will be obtained. 

Beyond the simple Gaussian context, more care is needed, since poor starting values can lead to poor tuning of non-conjugate samplers, and a consequent failure to properly explore the region of high posterior probability (along then with high sensitivity to the parameters of the smoothing parameter priors). \code{jagam} adopts the \pkg{mgcv} default smoothing parameter initializations and then performs one step of the penalized iteratively re-weighted least squares method for GAM fitting, in order to obtain starting values for the coefficients which are compatible with the initial smoothing parameters. The initial coefficients and corresponding standard errors are also used to set the scale of any required uninformative priors on the model coefficients: the prior standard deviation is set to 10 times the sum of the absolute value of the initial coefficient estimate and its standard error. 

\subsection{Further inference}

Having setup a GAM for use in \proglang{JAGS} and simulated from it, the user will typically want to visualize the smooths and predict from them. In addition some notion of the effective degrees of freedom of the smooth is useful. 

An obvious way to visualize the smooths is to draw curves from the posterior, and either compute appropriate pointwise quantiles in order to produce credible intervals, or to simply plot the curves. Examples are given below. Alternatively, smooths may be plotted with `two standard error bands', in the manner introduced in \citet{h&t90}. To this end it is only necessary to compute the mean coefficients from the simulation, to use in place of coefficient estimates, $\hat {\bm \beta}$, and to compute the observed covariance matrix of the simulated coefficients, ${\bf V}_\beta$, from which the standard error bands are readily computed. In fact $\hat {\bm \beta}$ and ${\bf V}_\beta$ also complete the preliminary \code{gam} object produced by \code{jagam} sufficiently for prediction using \code{predict.gam}. 

Finally some notion of the effective degrees of freedom of the model and its component smooths is useful. In a simple Gaussian additive model context a measure of the model effective degrees of freedom is $\text{tr}({\bf F})$ where ${\bf F} = ({\bf X}\ts{\bf X} + \sum_j \lambda_j {\bf S}_j)^{-1} {\bf X}\ts{\bf X}$. In the generalized additive model context ${\bf F} = ({\bf X}\ts{\bf WX} + \sum_j \lambda_j {\bf S}_j)^{-1} {\bf X}\ts{\bf WX}$, where $\bf W$ is the diagonal matrix of iteratively re-weighted least squares weights used in fitting. In the presence of random effects it is better to use ${\bf F} = {\bf V}_\beta {\bf X}\ts{\bf WX}/\phi$, in which $\bf W$ is the diagonal IRLS weight matrix with the random effects set to their posterior expectations and $\phi$ is the scale parameter or its estimate. When ${\bf V}_\beta$ is computed by simulation then this latter definition has the advantage of including a component for smoothing parameter uncertainty, however to compute it in practice requires that the expected value of the response, \code{mu}, be monitored during simulation.  See chapter 4 of \citet{wood2006igam} for further discussion.

Given $\bf F$, then the effective degrees of freedom of component smooths are obtained by summing the leading diagonal elements of $\bf F$ corresponding to the coefficients of the smooth concerned. Notice that under substantial modification of the \code{jagam} template model (involving modification of the response distribution, for example), the weighted versions of $\bf F$ may make no sense. It may then be better to fall back on the effective degrees of freedom that would have been computed if the model were a simple Gaussian additive model, or to use the estimate proposed by \cite{plummer.dic}.

\section{Examples}

As two simple examples consider the union wages example and the Sitka growth example from  \citet{crainiceanu2005}. Both datasets are available in the \pkg{SemiPar} \proglang{R} package \citep{semipar}. Loading the \proglang{JAGS} \code{glm} module, via \code{load.module("glm")} improves the efficiency of both examples in this section. 

\subsection{The union wages data}

The data frame `trade.union' contains a binary indicator of whether or not a worker is a trade union member, along with their hourly wage in US dollars. Consider the simple logistic regression model 
$$
\text{logit}(p_i) = f({\tt wage}_i), ~~~~ {\tt union.member}_i \sim \text{Bernoulli}(p_i)
$$
where smooth function $f$ is represented by a rank 20 thin plate regression spline. A \code{jagam} call sets the model up
\begin{Code}
R> jd <- jagam(union.member ~ s(wage, k=20), data = trade.union,
               family = binomial, file = "union.jags")
\end{Code}
resulting in the following \proglang{JAGS} model specification file.
\begin{Code}
model {
  eta <- X 
  for (i in 1:n) { mu[i] <-  ilogit(eta[i]) } ## expected response
  for (i in 1:n) { y[i] ~ dbin(mu[i],w[i]) } ## response 
  ## Parameteric effect priors CHECK tau is appropriate!
  for (i in 1:1) { b[i] ~ dnorm(0,0.018) }
  ## prior for s(wage)... 
  K1 <- S1[1:19,1:19] * lambda[1]  + S1[1:19,20:38] * lambda[2]
  b[2:20] ~ dmnorm(zero[2:20],K1) 
  ## smoothing parameter priors CHECK...
  for (i in 1:2) {
    lambda[i] ~ dgamma(.05,.005)
    rho[i] <- log(lambda[i])
  }
}
\end{Code}
The following commands then compile and simulate from the model. 
\begin{Code}
R> require(rjags); load.module("glm")
R> jm <- jags.model("union.jags", data = jd$jags.data,
R+                  inits = jd$jags.ini, n.chains = 1)
R> sam <- jags.samples(jm, c("b", "rho", "mu"), n.iter = 10000, thin = 10)
\end{Code}
On a 3GHz mid range laptop computer, simulation took 18 seconds, yielding effective sample sizes averaging around 400 for \code{rho} and 800 for \code{b} and \code{mu}, for the 1000 samples stored. \citet{crainiceanu2005} report around 9 minutes for this model (albeit with a slightly different smoothing penalty) for the same simulation length, on a 3.6GHz PC, although they do not report effective sample sizes, so the comparison is not completely straightforward. Note that failing to supply starting values greatly increases the adaptation and burn in time required to achieve reliable results for this example.   

\begin{figure}
\centering
\includegraphics[angle=-90,scale=0.65]{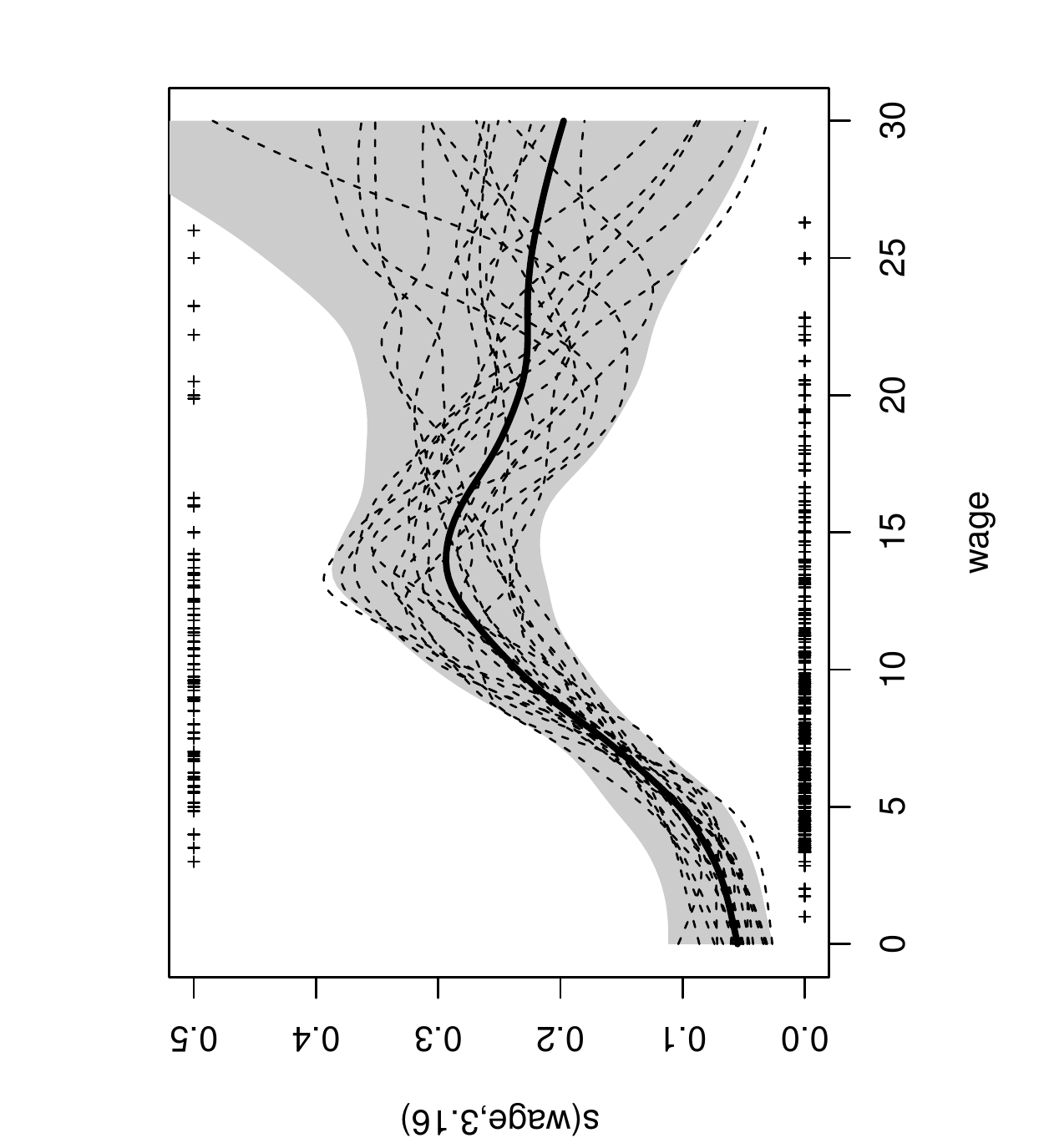}
\caption{Results for the Union Wages model. \label{union}}
\end{figure}

Finally a partial \code{gam} object can be created for convenient plotting and prediction. The following then produces a plot of the modelled probability of union membership against wages with a credible interval, along with a visualization of the union membership data. The interval is wide at high wages, failing to provide a very useful indication of the range of smooth shapes compatible with the data, so the following code also adds a sample of 20 curves from the posterior.  
\begin{Code}
R> jam <- sim2jam(sam, jd$pregam)
R> plot(jam, shade = TRUE, shift = coef(jam)[1], trans = binomial()$linkinv,
R+      rug = FALSE, ylim = c(-100, -coef(jam)[1]), seWithMean = TRUE,
R+      xlim = c(0, 30), lwd = 3)
\end{Code}
Given the basic plot, now add the original union membership data.
\begin{Code}
R> nu <- trade.union$union.member == 0
R> with(trade.union, points(wage[nu], 0 * wage[nu], pch = 3, cex = .5))
R> with(trade.union, points(wage[!nu], 0 * wage[!nu] + .5, pch = 3, cex = .5))
\end{Code}
And now add 20 smooth curves drawn from the posterior, to examine variability in the smooth shape.
\begin{Code}
R> ii <- 1:20 * 50; pd <- data.frame(wage = 0:300 / 10)
R> Xp <- predict(jam, type = "lpmatrix", newdata = pd)
R> for (i in ii) {
R+  p <- binomial()$linkinv(Xp 
R+  lines(pd$wage, p, lty = 2)
R+ }
\end{Code}
The result is shown in Figure \ref{union}, indicating that the peak in the probability curve is not a very robust feature, and it would be difficult to rule out a monotonic relationship between wages and probability of union membership.

\subsection{The Sitka growth data}

This example illustrates the  modification of an auto-generated \proglang{JAGS} model file to implement random effects. The `sitka' data contain repeated measurements over time of log size for Sitka spruce saplings grown under conditions of enhanced ozone, or control conditions.  A simple model has a smooth effect for time, a random intercept for each tree and an ozone effect,
$$
\log({\tt size}_i) = \alpha + f({\tt days}_i) + \beta {\tt ozone}_i + b_{j(i)} + \epsilon_i
$$
where $\epsilon_i \sim N(0,\sigma^2)$, $b_j \sim N(0,\sigma^2_b)$, and $j(i)$ is the index of the tree from which the $i^{\rm th}$ measurement is taken. Everything is Gaussian, so a fully conjugate setup can be employed, and it is worth diagonalizing the smoothing penalties.
\begin{Code}
R> jd <- jagam(log.size ~ s(days) + ozone, data = sitka,
R+             file = "sitka0.jags", diagonalize = TRUE)
\end{Code}

creates a default \proglang{JAGS} model file which can be modified to include the random effect as follows, where the blue italic code has been added to the non-italic auto-generated \proglang{JAGS} code (and auto-generated comments have been removed).

\begin{CodeChunk}
\begin{CodeOutput}
model {
  mu0 <- X %*% b 
\end{CodeOutput}
\vspace*{-.8cm}
\color{blue}
\begin{CodeInput}
  for (i in 1:n) { mu[i] <- mu0[i] + d[id[i]] } 
\end{CodeInput}
\vspace*{-.8cm}
\color{black}
\begin{CodeOutput}
  for (i in 1:n) { y[i] ~ dnorm(mu[i], tau) } 
  scale <- 1 / tau 
  tau ~ dgamma(.05, .005) 
\end{CodeOutput}
\vspace*{-.8cm}
\color{blue}
\begin{CodeInput}  
  for (i in 1:nd) { d[i] ~ dnorm(0, taud) }
  taud ~ dgamma(.05, .005)
\end{CodeInput}
\vspace*{-.8cm}
\color{black}
\begin{CodeOutput}
  for (i in 1:2) { b[i] ~ dnorm(0, 3e-04) }
  for (i in 3:10) { b[i] ~ dnorm(0, lambda[1]) }
  for (i in 11:11) { b[i] ~ dnorm(0, lambda[2]) }
  for (i in 1:2) {
    lambda[i] ~ dgamma(.05, .005)
    rho[i] <- log(lambda[i])
  }
}
\end{CodeOutput}
\end{CodeChunk} 

\begin{figure}
\centering
\includegraphics[angle=-90,scale=0.42]{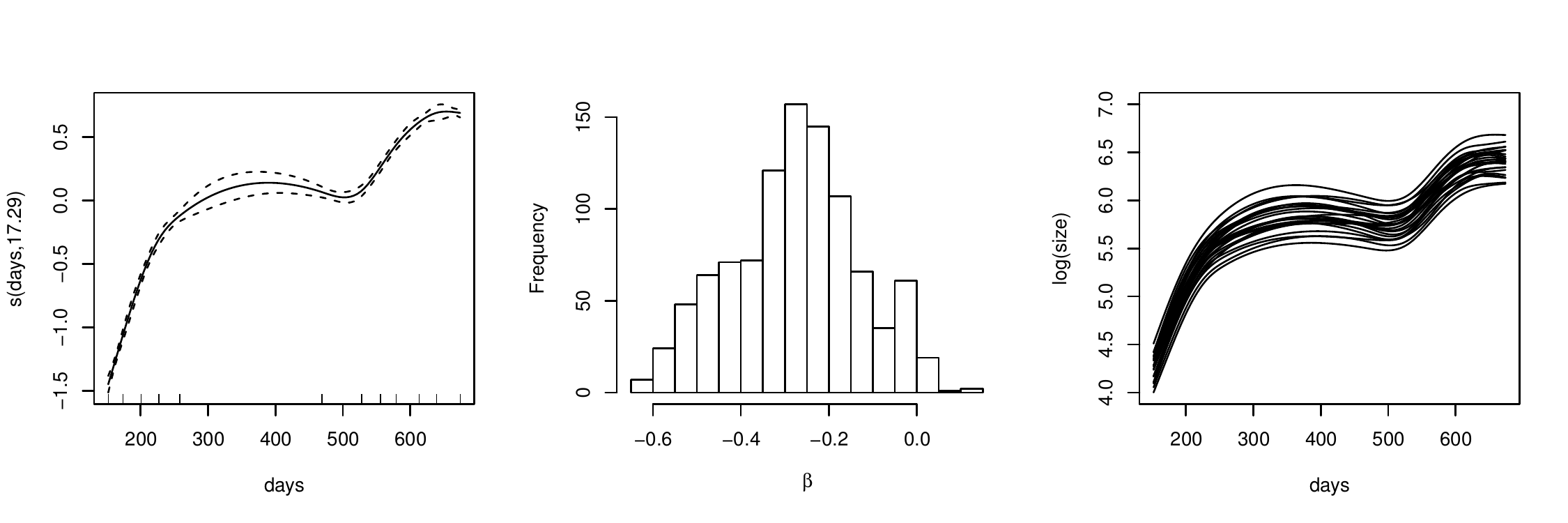}
\caption{Results for the Sitka growth model. \label{sitka}}
\end{figure}

 The modification involves adding a tree specific random effect, \verb+d+, to the linear predictor. Notice the need to add \code{id}, the vector attributing measurements to trees, and \code{nd}, the number of trees, to the \proglang{JAGS} data. The following code compiles and simulates from the model, produces a default plot of the smooth effect of time (given the sum to zero identifiability constraint), a histogram of draws from the posterior for $\beta$ and illustration of 25 curves, $\alpha + f({\tt days})$, drawn from the posterior.  
\begin{Code}
R> jd$jags.data$id <- sitka$id.num
R> jd$jags.data$nd <- length(unique(sitka$id.num))
R> jm <- jags.model("sitka.jags", data = jd$jags.data,
R+                 inits=jd$jags.ini, n.chains = 1)
R> sam <- jags.samples(jm, c("b", "rho", "scale", "mu"),
R+                     n.iter = 10000, thin = 10)
R> jam <- sim2jam(sam, jd$pregam)
R> plot(jam, pages = 1)
R> hist(sam$b[2, , 1])
R> days <- 152:674
R> pd <- data.frame(days = days, ozone = days * 0)
R> Xp <- predict(jam, newdata = pd, type = "lpmatrix")
R> ii <- 1:25 * 20 + 500
R> for (i in 1:25) {
R+  fv <- Xp 
R+  if (i==1) plot(days, fv, type = "l", ylim = c(4, 7)) else 
R+            lines(days, fv)
R+ }
\end{Code}
The results are shown in Figure \ref{sitka}. Notice how the left hand plot, which shows the credible interval for $f$ {\em subject to constraint}, suggests a very limited range of shapes for $f$. This is born out by the right hand plot, in which most of the variability in the curves is in their level, rather than their shape. 

\section{Conclusion}

The \proglang{JAGS} software offers enormous flexibility in the specification of complex random effects structures. Incorporating spline type smoothers into such models is routine, but somewhat tedious to code on a case by case basis, as well as being prone to error, especially for smooths of several variables. The \code{jagam} function offers a useful automation of the process of incorporating any smooth built into \pkg{mgcv} into a \proglang{JAGS} model, while dealing seamlessly with initialization and centring constraints and allowing straightforward posterior prediction. 

The main disadvantage of the approach is computational speed. Gibbs sampling for these models can be slow, especially if covariates are correlated. Indeed if only simple random effects are required then the random effects already available in \pkg{mgcv} may be much faster computationally. Similarly \pkg{BayesX} \citep[see e.g.,][]{fahrmeir.lang, fahrmeir04} is a substantially more efficient route to fully Bayesian inference with GAMs if the flexibility of \proglang{JAGS} is not required, while \proglang{Stan} (\url{http://mc-stan.org/}) offers another alternative likely to offer efficiency advantages. In these correlated settings it is also likely that Hamiltonian Monte Carlo methods \citep[e.g.,][]{girolami2011riemann} would enhance efficiency.

\subsection*{Acknowledgements}

I am grateful to Ciprian Crainiceanu and Mirjam Barrueto for some very helpful discussion of \cite{crainiceanu2005}, and to two referees for pointing out how much the \code{glm} library can improve performance, suggestions on the `fixed effect' priors and other useful comments. This work was funded by UK EPSRC grant EP/K005251/1.

\bibliography{jss1488j,jss1488}

\end{document}